\documentclass[12pt,epsf]{article}
\usepackage{graphicx,float,subfigure,epsfig,color,rotating,latexsym}
\usepackage[latin1]{inputenc}
\usepackage[T1]{fontenc}
\def\be{\begin{equation}}
\def\ee{\end{equation}} 
\def\bea{\begin{eqnarray}}
\def\eea{\end{eqnarray}} 
\def\ba{\begin{array}} 
\def\ea{\end{array}}

\def\la{\lambda} 
\def\La{\Lambda}

\def\nin{\noindent} 

\def\ket{\rangle}
\def\bra{\langle}

\begin{document}

\begin{center} {\Large{\bf Effective potential and vacuum stability }}\\

\vspace*{0.7 cm}
Vincenzo Branchina\footnote{vincenzo.branchina@ct.infn.it}\label{one} 
\vspace*{0.2 cm}

{\it Department of Physics, University of
Catania and \\ INFN, Sezione di Catania, 
Via Santa Sofia 64, I-95123, Catania, Italy }

\vspace*{0.5 cm}

Hugo Faivre\footnote{hugo.faivre@ires.in2p3.fr}\label{two},\,
Vincent Pangon\footnote{vincent.pangon@ires.in2p3.fr}\label{three}
\vspace*{0.2cm}

{\it IPHC/DRS, Theory Group - Louis Pasteur University and CNRS,} \\
\vspace*{0.02cm}
{\it 23 rue du Loess, 67037 Strasbourg, France}\\

\vspace*{1 cm}

{\LARGE Abstract}\\

\end{center}

\vspace*{0.1cm}

\noindent
By following previous work on this subject, we investigate the 
issue of the instability of the electroweak vacuum against the 
top loop corrections by performing an accurate analysis of a 
Higgs-Yukawa model. We find that, when the physical cutoff is 
properly implemented in the theory, the potential does not 
exhibit any instability. 
Moreover, contrary to recent claims, we show that 
this instability cannot be understood in terms of the very 
insightful work of Wu and Weinberg on the non-convexity of the 
one-loop effective potential of a scalar theory. Some of the 
theoretical and phenomenological consequences of our results 
are briefly discussed.

\vspace*{0.5cm}
\section{Introduction}

It is commonly believed that the top loop contribution to the 
Higgs effective potential $V_{eff}(\phi)$ 
destabilizes the electroweak (EW) vacuum \cite{list}. According 
to our recent analysis, however, this instability 
results from extrapolating $V_{eff}(\phi)$ beyond its region 
of validity \cite{us}. When we limit ourselves to the range 
of $\phi$ where the approximations considered for its computation 
hold, $V_{eff}(\phi)$ turns out to be a convex (therefore, a 
fortiori, stable) function of its argument, in agreement 
with well known exact theorems \cite{sim,iliop}. 

In order to avoid unnecessary complications related to the 
gauge sector of the Standard Model (SM), in \cite{us} we started
our analysis by considering a Higgs--Yukawa model. 
In fact, for large values of the Yukawa coupling, this model 
presents the same (apparent) instability of the SM. 

Our work has been recently challenged in \cite{einh} and this 
provides one of the motivations for the present paper. By  
pursuing our investigation on this issue, we shall see  
that some of the arguments considered in \cite{einh} are 
incorrect, while some others do not apply to the problem under 
investigation. The results of our present analysis, in fact,  
reinforce the conclusions of our previous work: the vacuum 
instability is unphysical and is nothing but the result of an 
illegal extrapolation of the computed potential beyond its 
region of validity. 

Let us consider the Higgs--Yukawa model: 

\be\label{lagra}
{\mathcal L}(\phi,\psi,\overline\psi)= 
\frac12\partial_\mu\phi\partial_\mu\phi 
+\overline\psi\gamma_\mu\partial_\mu\psi+\frac{m_{_\La}^2}{2}\phi^2
+\frac{\la_{_\La}}{24}\phi^4 + g_{_\La}\phi\overline\psi\psi \,,
\ee 

\nin
where $\La$ is the ultraviolet cutoff of the theory and 
$m^2_{_\La}$, $\la_{_\La}$ and $g_{_\La}$ 
are the bare mass and coupling constants respectively. 
The bare potential for the scalar field is:

\bea\label{potcl}
V_{_\La}(\phi)&=&\frac{m^2_{_\La}}{2}\phi^2+\frac{\la_{_\La}}{24}\phi^4 \,.
\eea 
\nin
A straightforward computation of the one-loop effective 
potential\footnote{We use the notation 
$V^{(Y)}_{1l}(\phi)$ to indicate the Higgs-Yukawa one-loop effective 
potential, while the one-loop potential of the scalar theory alone 
will be indicated in the following with $V^{(s)}_{1l}(\phi)$.}  
$V^{(Y)}_{1l}(\phi)$ gives the well known result (see, for instance, \cite{us}):  
\bea\label{Vdimreg}
V^{(Y)}_{1l}(\phi)=\frac{m^2}2\phi^2+\frac{\la}{24}\phi^4
+\frac{\left(m^2+\frac{\lambda}{2}\phi^2\right)^2}{64\pi^2}\left({\rm ln}
\left(\frac{m^2+\frac{\lambda}{2}\phi^2}{\mu^2}\right)-\frac{3}{2}\right)
-\frac{g^4\phi^4}{16\pi^2}\left({\rm ln} \frac{g^2\phi^2}{\mu^2} 
-\frac{3}{2}\right) ,
\eea
where $m^2\equiv m^2_\mu$, $\la\equiv \la_\mu$, $g\equiv g_\mu$ and $\mu$ 
is the renormalization scale.

With the help of renormalization group (RG) techniques, it is possible 
to improve on this result \cite{jones}. For the purposes of 
our analysis, however, we can limit ourselves to consider the one-loop 
approximation to $V_{eff}(\phi)$, as the RG-improvement has no relevance  
to our discussion (see \cite{us} for details). 

We see from Eq.(\ref{Vdimreg}) that, when the 
Yukawa coupling is sufficiently strong, $V^{(Y)}_{1l}(\phi)$ 
becomes unstable for large values of $\phi$. Although this is 
at odds with the convexity property of the exact 
$V_{eff}(\phi)$, it is argued in \cite{einh} that  
$V^{(Y)}_{1l}(\phi)$ can be trusted even in the region 
where it is not convex. 
The authors base their reasoning on the analysis presented in 
\cite{wuwein} (see also \cite{piguth}), where the issues of 
the physical meaning of the imaginary part of the one-loop 
effective potential $V^{(s)}_{1l}(\phi)$ of the self-interacting 
scalar theory alone (i.e. the theory obtained from Eq.(\ref{lagra}) 
for a vanishing Yukawa coupling) and the 
non-convexity of its real part were addressed. 

As we shall explain in the 
following (see section 2), although physically meaningful and 
relevant to the issues considered in that paper, the analysis 
put forward in \cite{wuwein} bears no relation with the 
problem of the instability induced by the fermion loop 
contribution to $V^{(Y)}_{1l}(\phi)$. In fact, it turns out 
that the origin of the non-convexity of $V^{(Y)}_{1l}(\phi)$ 
in the region of large $\phi$, which is our problem, is {\it totally 
different} from the origin of the non-convexity
of $V^{(s)}_{1l}(\phi)$ in the region between the classical minima,
which is the problem studied in \cite{wuwein}. The interpretation 
found in \cite{wuwein} for the non convex part of $V^{(s)}_{1l}(\phi)$
{\it cannot be extended} to the non-convexity of  
$V^{(Y)}_{1l}(\phi)$. 

The paper is organized as follows. In section 2, we begin by briefly 
reviewing the analysis of \cite {wuwein} on the interpretation 
of the non-convex and non-real parts of the one-loop approximation 
to the effective potential $V^{(s)}_{1l}(\phi)$ of the scalar 
self-interacting theory. 
Then, we show why, contrary to recent claims \cite{einh}, these 
results bear no relation with the instability shown by the 
Higgs-Yukawa one-loop potential $V^{(Y)}_{1l}(\phi)$. In section 3, 
we show that the 
region of $\phi$ where the renormalized Higgs-Yukawa one-loop 
potential apparently becomes unstable lies beyond the region of
validity of perturbation theory. In section 4, we draw our 
conclusions.

\section{Instability in the internal and in the external regions}

For the purposes of our analysis, it is worth to recall the 
following definition of the effective potential 
$V_{eff}(\phi)$ for a single component scalar theory. 
Given the Hamiltonian density $\hat \mathcal H$, 
$V_{eff}(\phi)$ is obtained by minimizing the expectation value of 
$\hat \mathcal H$, 
\be\label{veff}
V_{eff}(\phi)= min\bra \Psi|\hat \mathcal H|\Psi\ket \,,
\ee
with the constraints,
\be\label{constr}
\bra \Psi|\Psi\ket =1 \mbox {~~~~~~~ and ~~~~~~~ }  \bra \Psi|\hat \phi|\Psi\ket =\phi\,.
\ee 

In order to support the reliability of the instability 
induced by the fermion loop contribution to $V^{(Y)}_{1l}(\phi)$, 
the authors of \cite{einh} invoke the work of Wu and Weinberg \cite{wuwein}. 
However, we now show that the analysis presented
in \cite{wuwein}, although very illuminating 
on the physical meaning of $V^{(s)}_{1l}(\phi)$ (the one-loop effective 
potential of a single component scalar self-interacting theory), 
has nothing to do with the instability problem under investigation. 
To this end, let us first briefly review 
the main results of \cite{wuwein}. 

\begin{figure}[ht]
\begin{center}
\includegraphics[width=7cm,height=11cm,angle=270]{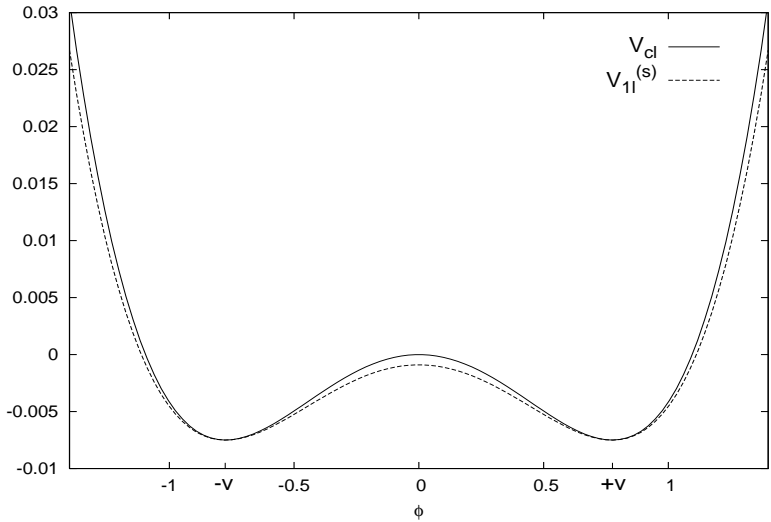}
\end{center}
\caption[]{The continuous line represents the classical potential
$V_{cl}(\phi)$ of a single component scalar theory in the 
broken phase, while the dashed line is the real part of the one-loop 
effective potential $V^{(s)}_{1l}(\phi)$. The latter provides a good 
approximation for the 
real part of $\widetilde V(\phi)$, the modified effective potential 
for localized states (see text).}
\label{fig1}
\end{figure}

In this work, the authors consider a single component scalar 
theory which exhibits symmetry breaking at the classical level. 
As is well known, $V^{(s)}_{1l}(\phi)$ develops an imaginary
part for values of $\phi$ between the inflection
points of the potential. The aim of \cite{wuwein} is to investigate
the physical meaning of the real and imaginary parts of $V^{(s)}_{1l}(\phi)$. 
After a thorough analysis, they come to 
the conclusion that the function that $V^{(s)}_{1l}(\phi)$ is approximating
{\it is not} $V_{eff}(\phi)$ as defined by Eqs.(\ref{veff}) and (\ref{constr}),
but a sort of {\it modified effective potential},
which they call $\widetilde V(\phi)$, 
obtained by requiring that the state $|\Psi\ket $ (i.e. the wave functional 
$\Psi[\phi]=\bra \phi|\Psi\ket $) be {\it localised} \footnote{This question 
was already discussed by P.M. Stevenson in \cite{stev} in the framework of 
the Gaussian approximation to the effective potential.} 
around $\phi$. The real part of $V_{1l}(\phi) \simeq \tilde V(\phi)$ 
(see Fig.\ref{fig1}) turns out to be  the expectation 
value of the energy density on such a localized (unstable) state while its 
imaginary part is half its decay rate per unit volume \cite{wuwein}. 

\begin{figure}[ht]
\begin{center}
\includegraphics[width=7cm,height=11cm,angle=270]{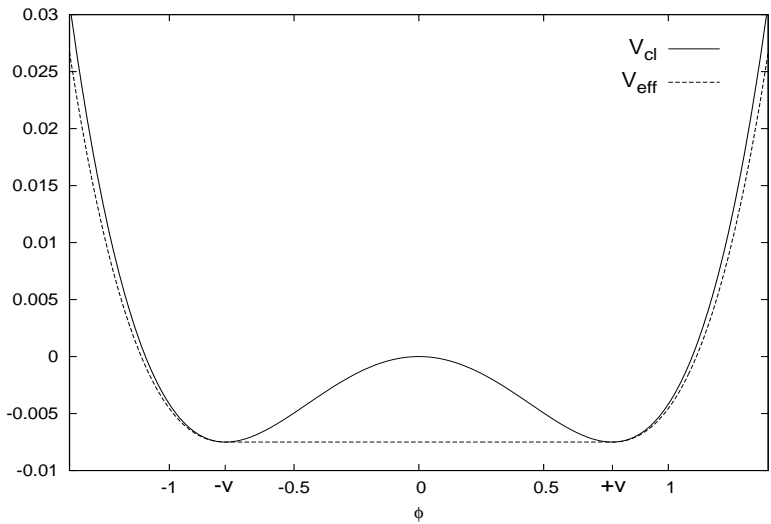}
\end{center}
\caption[]{The continous line represents the classical potential
$V_{cl}(\phi)$ of a single component scalar theory
in the broken phase, while the dashed line is the effective potential
$V_{eff}(\phi)$. In the internal region (i.e. in the region 
between the classical minima), $V_{eff}(\phi)$ is well approximated 
by the Maxwell construction (internal dashed line). In the external 
region it is well approximated by $V^{(s)}_{1l}(\phi)$ (external 
dashed line). As is clear from the figure, this 
approximation of $V_{eff}(\phi)$ is a convex function of $\phi$.}
\label{fig2}
\end{figure}

As for $V_{eff}(\phi)$, the true effective potential, it turns out that 
in the region between the classical minima $-v $ and $+v $ (the so called 
{\it internal region}) it is well approximated by the Maxwell construction, 
whereas in the region where $|\phi| \geq v$ (the {\it external region})
$V_{1l}(\phi)$ provides a good approximation to 
$V_{eff}(\phi)$ \cite{wuwein, rivers1} (see Fig.\ref{fig2}). 
Actually, for any given value of $\phi$ in the internal region, 
the true vacuum state of the system is given by an inhomogeneous mixed 
state which is a linear combination of $ |+v \ket $ and $ |-v \ket $
\cite{wuwein}.

Within the framework of the loop expansion, the origin of the Maxwell 
construction can be easily understood once we realize that in the 
internal region the one-loop approximation, which is based on the 
tacit assumption that the 
path integral is dominated by a single saddle point,  
looses its validity. For those values of $\phi$, there are {\it two} 
competing non trivial saddle points which contribute with the same weight
to the path integral. Taking into account the contribution 
of both of them, the flat shape of $V_{eff}(\phi)$ between $-v $ and 
$+v $ (Fig.\ref{fig2}) immediately arises \cite{rivers1}. 
This also explains 
why, in the external region, the usual loop expansion, in particular 
the one-loop approximation $V_{1l}(\phi)$, provides a good approximation 
to $V_{eff}(\phi)$. In fact, for those values of $\phi$, there is no 
competition of saddle points; the path integral is saturated by a single 
saddle point and the usual loop expansion is at work. 

\begin{figure}[ht]
\begin{center}
\includegraphics[width=7cm,height=11cm,angle=270]{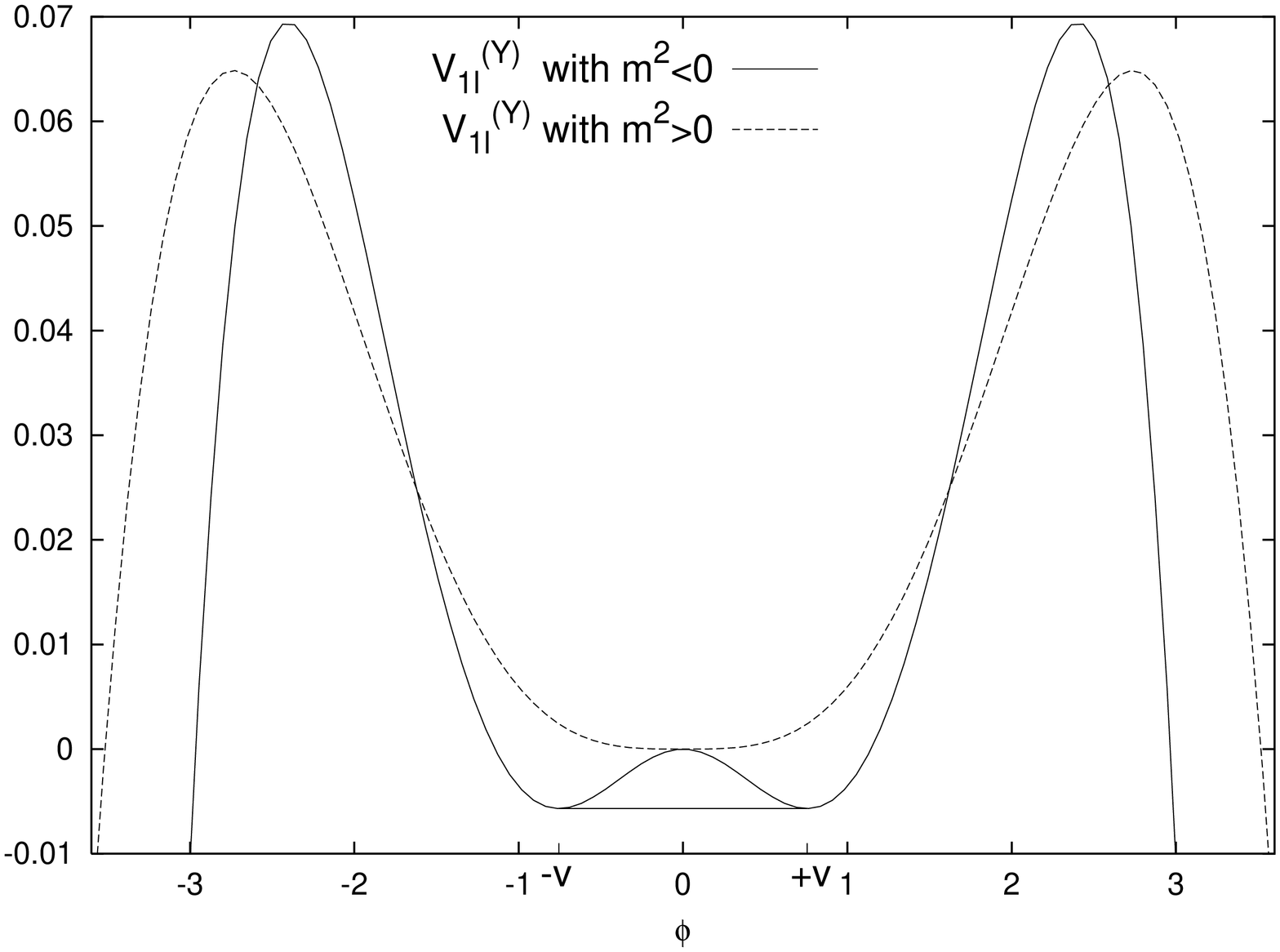}
\end{center}
\caption[]{Continous line: the renormalized one-loop potential 
$V^{(Y)}_{1l}(\phi)$ of the Higgs-Yukawa model as given in 
Eq.(\ref{Vdimreg}) ($m^2 <0$). The (apparent) instability occurs in the 
external region, i.e. for $ |\phi| > + v $, far away from the 
internal region where the competition of two saddle points 
with the same weight gives rise to the Maxwell construction,
the horizontal line between $- v$ and $+v$  
(see text). Dotted line: $V^{(Y)}_{1l}(\phi)$ for a positive 
value of $m^2$. As noted in the text, while in the internal 
region the one-loop potential is now convex, the apparent instability 
in the external region still persists. This figure clearly 
shows the different origin of the non-convexity in the internal 
and external regions.}
\label{fig3}
\end{figure}  
 
Having reviewed the main results of \cite{wuwein}, now we can see 
why this work (although interesting for the insight it provides 
on the physical understanding of $V^{(s)}_{1l}(\phi)$) bears 
no relation with 
(therefore, cannot be invoked to support the reliability of) 
the instability induced by the fermion loop corrections in the 
one-loop effective potential $V^{(Y)}_{1l}(\phi)$ of the 
Higgs-Yukawa model. 

We begin by noting that this instability (non-convexity) 
occurs in the external region (see Fig.\ref{fig3}), the 
region beyond the minima, while the analysis of \cite{wuwein} 
concerns the non-convexity of the effective potential in the internal 
region (see Fig.\ref{fig2}), the region between the minima. 
Moreover, the instability in our problem results from the integration over the 
fermion modes and this has nothing to do with the internal region 
non-convexity studied in \cite{wuwein}, which results from the 
competition of two non trivial saddle points with the same weight. 
It is then clear that the external region non-convexity 
of $V^{(Y)}_{1l}(\phi)$ has a {\it totally different origin} from 
the internal region non-convexity. 

The fact that, for sufficiently large values 
of the field, $V^{(Y)}_{1l}(\phi)$ bends down is due to the 
negative contribution of the fermion loop, the term
$ - g^4\phi^4 \,{\rm ln} \,\phi^2$ of Eq.(\ref{Vdimreg}), 
which overwhelms the classical $\lambda \phi^4$ term for large 
values of $\phi$. Obviously, this has nothing to do with the 
non-convexity in the internal region, which 
comes from the presence of a negative $m^2$. In this respect,
it is worth to note that the external region instability of 
$V^{(Y)}_{1l}(\phi)$ occurs even for positive values of $m^2$
(see Fig.3). 

We are now in the position to draw our first conclusion. 
Although at this stage of our analysis we do not know yet 
whether the instability of $V^{(Y)}_{1l}(\phi)$ is a physical
effect or not, we have already established that   
(contrary to what is claimed in \cite{einh}) it is not 
possible to invoke the fact that the non-convexity of 
$V^{(s)}_{1l}(\phi)$ in the internal region has a sensible 
physical interpretation, which is the result of \cite{wuwein}, to 
argue that the instability of $V^{(Y)}_{1l}(\phi)$ in the external 
region is a genuine physical effect. 

Therefore, the physically relevant question to ask at this point 
is the following. 
Can we give a physically meaningful interpretation to the non-convex 
part of $V^{(Y)}_{1l}(\phi)$ in the external region (Fig.\ref{fig3}) 
as we can do for the non-convexity in the internal one \cite{wuwein}?
In the following section we shall find that, in this region, 
the approximations used for the computation of the renormalized 
$V^{(Y)}_{1l}(\phi)$ break down. Moreover, we shall see that, when 
the cutoff is properly implemented in the theory, no instability 
occurs.
 
\section{Instability within perturbation theory and RG}

Going back to the renormalized Higgs-Yukawa 
potential $V^{(Y)}_{1l}(\phi)$ of Eq.(\ref{Vdimreg}), we 
see that the instability is driven by the fermion contribution, 
which is also the dominant correction term. Therefore, 
with no loss of generality, we can neglect in this
equation the boson contribution, as well as other unimportant finite 
terms, so that the one-loop potential takes the (simplified) form:

\be\label{Vnew}
V^{(Y)}_{1l}(\phi)\simeq \frac{m^2_\mu}{2}\phi^2 + \frac{\la_\mu}{24}\phi^4
-\frac{g^4\phi^4}{16\pi^2}\,{\rm ln} \frac{\phi^2}{\mu^2} \, .
\ee
The first two terms correspond to the classical potential 
$V_{cl}(\phi)$, which is written in terms of the IR 
parameters $m^2_\mu$ and $\la_\mu$, with 
$m^2_\mu <0$ and $\la_\mu > 0$.  
The potential is typically normalized so that 
\be \label{zero}
V^{(Y)}_{1l} (\pm v)=0\, , 
\ee
where $\pm v$ are the classical minima. 
Then, the instability takes place when 
$V^{(Y)}_{1l}(\phi)$ becomes negative. Note also 
that, as the instability occurs 
for large values of $\phi$, in the following discussion we can 
further simplify Eq.(\ref{Vnew}) by neglecting the mass term and
write:   
\be\label{Vnew2}
V^{(Y)}_{1l}(\phi)\simeq\, \frac{\la_\mu}{24}\phi^4
-\frac{g^4\phi^4}{16\pi^2}\,{\rm ln} \frac{\phi^2}{\mu^2} \, .
\ee

A point which is relevant to our analysis concerns the 
sign of $\lambda_{_{\Lambda}}$. In \cite{einh} is said that 
there is no reason why it should be positive, and we would like 
to turn our attention to this issue. 
First we note that, with the help of simple renormalization 
group analysis, the effective potential in 
Eq.(\ref{Vnew2}) can be written as: 
\be\label{Vnew3}
V^{(Y)}_{1l}(\phi)\simeq\, \frac{\la_\mu}{24}\phi^4
-\frac{g^4\phi^4}{16\pi^2}\,{\rm ln} \frac{\phi^2}{\mu^2} 
=\frac{\la_{_{\Lambda}}}{24}\phi^4
+\frac{g^4\phi^4}{16\pi^2}\,{\rm ln} \frac{\Lambda^2}{\phi^2}\, .
\ee
Eq.(\ref{Vnew3}) tells us 
that, if $\lambda_{_{\Lambda}}$ is positive, $V^{(Y)}_{1l}(\phi)$ 
is also positive, at least in the region where 
it is defined, i.e. for 
$\phi < \Lambda$. Therefore, from Eq.(\ref{zero}) we see that no 
instability can occur. 

Now, it is not difficult to convince ourselves that 
$\lambda_{_{\La}}$ is positive. From a physical point of 
view, in fact, the theory has to be regarded as an effective 
theory, intrinsically defined with a cutoff $\La$,  
the higher energy scale where it is still valid. The bare 
potential, which defines the theory at the scale $\La$, 
has to be well behaved. For a negative 
value of $\lambda_{_{\La}}$, however, we would have a potential 
unbounded from below and the theory would not be defined. 

Moreover, we observe that, within the framework of cutoff perturbation 
theory, in order to get the 
$ - g^4\phi^4\,{\rm ln}({\phi^2}/{\mu^2})$ term, which is responsible 
for the instability of 
$V^{(Y)}_{1l}(\phi)$ in Eq.(\ref{Vdimreg}), we have to consider 
the counterterm \, $\frac{\delta\la}{24}\,\phi^4$,\, where: 
\be\label{deltalam}
\frac{\delta\la}{24} =
-\frac{g^4}{16\,\pi^2}\,{\rm ln} \frac{\La^2}{\mu^2} \,.
\ee
Let us assume now that a range of $\phi$'s exists such that 
$V^{(Y)}_{1l}(\phi) <0$ and that in this region 
$V^{(Y)}_{1l}(\phi)$ can be trusted (so that we can claim instability
of the potential). As:  
\be\label{cut}
\Lambda \,>\, \phi \,,
\ee
\nin
we immediately have $V^{(Y)}_{1l}(\La)<V^{(Y)}_{1l}(\phi)<0$, which 
combined with Eq.(\ref{Vnew3}) gives:
\be
\frac{\la_\La}{24} 
= \frac{\la_\mu}{24} 
-\frac{g^4}{16\,\pi^2}\,{\rm ln} \frac{\La^2}{\mu^2} \,  < \, 0 \,. 
\ee 
Being the renormalized quartic coupling constant $\la_\mu$ positive, 
which is a necessary condition to have a non trivial classical 
minimum, the result $\la_\La <0$ implies that (the absolute value of) 
$\delta \la$ must be greater than $\la_\mu$ itself, 
which would signal a breakdown of perturbation theory. 

Finally, we note that the result 
$\lambda_\Lambda = \lambda_\mu + \delta\la$, with $\delta\la$
given by Eq.(\ref{deltalam}), is nothing but the result that 
would be obtained with the help of the perturbative RG equation 
for the running coupling constant $\lambda (p)$ at $p=\Lambda$. 
Therefore, it is easy to rephrase the above considerations in a 
renormalization group language. The apparent instability of the 
effective potential is related to the fact that the RG equations 
would drive $\lambda_\mu$ to negative values in the high energy 
regime (see, for instance, \cite{kuti}). As we 
have just seen, however, this range of energies is not allowed. 

In this respect, we note that an interesting 
discussion, which closely parallels and complements 
our results, can be found in \cite{kuti}. 
In particular, the authors show that the renormalized 
perturbative RG equation for the quartic coupling constant 
incorrectly predicts a flow toward negative values of $\la$,
which in turn signals an (apparent) instability of the 
effective potential. However, they also show that, when the 
finite cutoff is correctly enforced in the effective Higgs-Yukawa 
theory, in the high energy regime the true flow diverges from the 
renormalized one. The coupling constant $\la$ do not turn to negative 
values and no instability occurs.  

\begin{figure}[ht]
\begin{center} 
\includegraphics[width=7cm,height=11cm,angle=270]{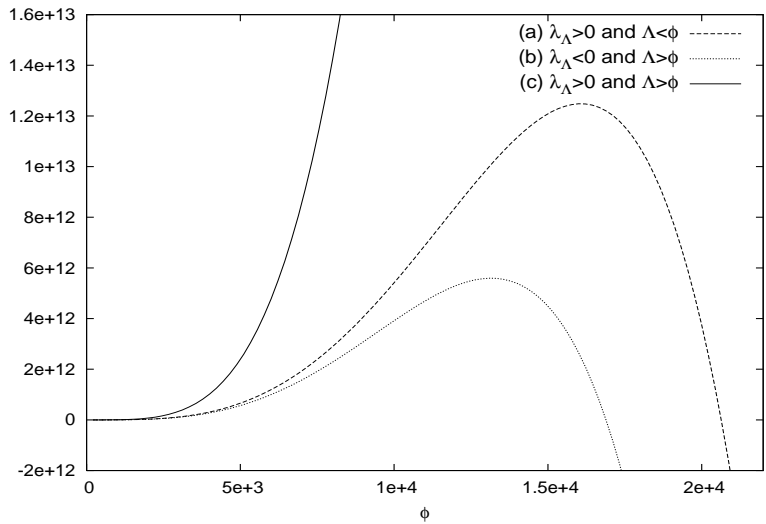}
\end{center}
\caption[]{The effective potential $V^{(Y)}_{1l}(\phi)$ for 
the three following cases (arbitrary units): 
(a) $\La=5\cdot 10^2$, $m^2_{\La}=-5\cdot 10^2$,  $\la_\La=5\cdot10^{-2}$. 
The potential becomes unstable for values of the field $\phi > \Lambda$ ; 
(b) $\La=2\cdot 10^4$, $m^2_{\La}=-5\cdot 10^2$, $\la_\La=- 5\cdot10^{-2}$.  
The potential becomes unstable for values of the field 
$\phi < \Lambda$. However, being $\la_{\La}<0$, the bare potential 
is unbounded from below and the theory is not defined.  
(c) $\La=2\cdot 10^4$, $m^2_{\La}=-5\cdot 10^2$, $\la_\La=5\cdot10^{-2}$. 
The potential is convex in the external region. In the internal 
region, as we know, the saddle point competition produces a flat 
potential (Maxwell construction).}
\label{figlast}
\end{figure}

Fig.\ref{figlast} contains an useful overview of our 
previous results. When the theory at the scale $\La$ is properly 
defined, i.e. when $\la_\La > 0$, and  $V^{(Y)}_{1l}(\phi)$ is 
considered only for values of the field such that $\phi < \La$, 
no instability occurs.

Before ending this section, it is worth to mention an  
additional point. In \cite{einh} is acknowledged 
that ``the perturbation expansion for V breaks down'' 
at the point where the non-convexity takes place, that is 
at the inflection points (see Fig.5 of \cite{einh}), which, 
in their notations, occurs when the ``renormalization group time'' 
$t={\rm ln}\, \mu$ takes the value $t\sim 2.98$. 
However, they note 
that for larger values of $t$ their perturbative parameter 
become again small, so they claim that perturbation theory 
results can be trusted. From the above analysis (as well as from 
the results of \cite{us} and \cite{kuti}), however, is clear 
that once at a given scale the results of the renormalized 
perturbation theory start to diverge from the effective theory (bare)
ones, this divergence is kept at all higher scales, no matter how 
small the perturbative parameter is. In other words, the region of 
instability of $V^{(Y)}_{1l}(\phi)$ lies well beyond the range of 
validity of the renormalized perturbative result for 
$V^{(Y)}_{1l}(\phi)$ itself. 

\section{Summary and conclusions}

In the present work we have shown that, contrary to recent 
claims \cite{einh}, the instability of the Higgs-Yukawa 
one-loop potential $V^{(Y)}_{1l}(\phi)$ 
induced by the fermion loop contribution {\it cannot} 
be understood in terms of the very insightful work of Weinberg and 
Wu \cite{wuwein}, where the physical meaning of the non-convex
part of the one-loop scalar effective potential 
$V^{(s)}_{1l}(\phi)$ (as well as of its imaginary part) was investigated. 
We have shown in some detail that the instability (non-convexity) of the 
Higgs-Yukawa potential $V^{(Y)}_{1l}(\phi)$ has a {\it completely different} 
origin from the non-convexity of $V^{(s)}_{1l}(\phi)$ in the 
region between the classical minima. 

Contrary to this latter case, where a physical meaning to the non-convex 
$V^{(s)}_{1l}(\phi)$ can be given \cite{wuwein}, the instability 
induced by the fermion loop in the Higgs-Yukawa potential results from an 
illegal extrapolation of $V^{(Y)}_{1l}(\phi)$ itself beyond its region of 
validity. From a renormalization group point of view, the instability 
comes from the fact that, for high energy scales, the RG equations 
{\it apparently} drive the quartic coupling constant $\lambda$ to negative 
values (see, for instance, \cite{kuti}, \cite{einh}). As we have shown, 
however, the perturbative RG equations are no 
longer valid at these scales (see also \cite{kuti}).

In this respect, it is worth to say a word of caution about
the dimensional regularization scheme. This scheme, which is very 
powerful to get the finite results of renormalized perturbation 
theory, can be safely used only when the latter holds. One of the 
most striking drawbacks of the method, which has some resemblance 
with our problem, concerns the Appelquist-Carazzone decoupling
theorem \cite{cara}. Being mass-independent, the $\overline {MS}$ 
scheme cannot reproduce the results of this theorem 
(see, for instance, \cite{mano}). 
Another, largely unnoticed, drawback concerns the instability 
problem. As we have seen, the instability of the effective 
potential occurs for values of $\phi$ beyond the physical cutoff.  
This can be detected only when a more physical scheme, as a momentum 
cutoff, is adopted. When this is done, the illusory nature of the 
effective potential instability becomes immediately clear 
(see also \cite{us} and \cite{kuti}).

Finally, it is worth to note that the instability problem is of 
theoretical as well as of phenomenological interest. On the theoretical 
side, we believe that our analysis shows that, in general, fermion
corrections as the ones occurring in the one-loop potential of the 
Higgs-Yukawa model, the top loop in the case of the SM, {\it cannot}
destabilize the classical vacuum. On the phenomenological side, it is 
well known that the most important application concerns the lower 
bounds on the Higgs boson mass. Our analysis shows that the traditional 
way of setting these bounds does not work. In \cite{us} and \cite{kuti}, 
however, novel methods for extracting lower bounds for the Higgs mass 
have been put forward.

\end{document}